\documentstyle[emulateapj]{article}

\submitted{Accepted for publication in the Astrophysical Journal, Letters}

\lefthead{Yasuhiro Shioya, Neil Trentham \& Yoshiaki Taniguchi}
\righthead{Arp 220}

\begin{document}

\title{ON THE HIDDEN NUCLEAR STARBURST IN ARP 220} 

\author{Yasuhiro Shioya$^1$, Neil Trentham$^2$, \& Yoshiaki Taniguchi$^{1,3}$}

\vspace {1cm}

\affil{$^1$Astronomical Institute, Graduate School of Science, 
       Tohoku University, Aramaki, Aoba, Sendai 980-8578, Japan}
\affil{$^2$Institute of Astronomy, University of Cambridge, Madingley Road, Cambridge
       CB3 0HA, UK}
\affil{$^3$Institute for Astronomy, University of Hawaii, 2680
 Woodlawn Drive, Honolulu, HI 96822}

\begin{abstract}
We construct a starburst model for the hidden starbursts
in Arp 220 based on the new Starburst99 models of Leitherer et al. 
Comparing these stellar population synthesis models with observations, 
we show that the  hidden power source must be due to star
formation (as opposed to an AGN) at the 50\% level or more in order to avoid an ionizing
photon excess problem, and this starburst must be young
($< 7 \times 10^7$ yr).
We derive a current star formation rate of
$270 M_{\odot}$ yr$^{-1}$, and an extinction $A_V > 30$ mag 
for our line of sight to this
hidden starburst.
\end{abstract}


\section{INTRODUCTION}

Ultraluminous infrared galaxies (ULIGs) are the most luminous 
galaxies in the Universe and their bolometric luminosities 
are comparable to those of quasars ($L_{\rm bol} > 10^{12}L_{\odot}$). 
The physical origin of the large luminosity of ULIGs 
is unknown since the very centers of these objects, which is
where the bolometric luminosity comes from (Sanders \& Mirabel 1996) is
very heavily obscured by dust and direct observation is not possible.
Either an enshrouded starburst or an 
active galactic nucleus (AGN), or both, could generate the
observed luminoisites.

Arp 220 is the nearest and best-studied 
ULIG\footnote{We use a distance of 74 Mpc, assuming $H_0=75 \; {\rm km \; s^{-1} \; Mpc^{-1}}$ and $\Omega_0=1$
with the recession velocity to the Galactic Standard of Rest,
$V_{\rm GSR}=5,531 {\rm km \; s^{-1}}$
(de Vaucouleurs et al. 1991).} 
Recently Genzel et al.~(1998) performed a mid-infrared spectroscopy
program for a sample of ULIGs and demonstrated that
Arp 220 has an appreciable PAH feature and so
is powered by a starburst, assuming that the very core where $L_{\rm bol
}$
originates from is not optically thick at
15 $\mu$m.
Based on detailed radiative transfer modelling, Rowan-Robinson 
(2000 and references therein) 
argues that galaxies like Arp 220 must be star-formation
powered since they emit most of their luminosity longward of 50$\mu$m. 
Additional indirect evidence that Arp 220 is powered by star formation
includes
\vskip 1pt \noindent
(i) the lack of an X-ray detection (Iwasawa 1999), even at hard wavelengths
where Compton scattering is less important;
\vskip 1pt \noindent
(ii) VLBI radio observations, which show many radio supernovae but
no strong point radio source which could correspond to the AGN
(Smith et al.~1998);
\vskip 1pt \noindent
(iii) a LINER\footnote{LINER = Low Ionization Nuclear
Emission-line Region (Heckman 1980).}-like optical emission-line spectrum
which can result from a starburst-driven superwind 
(Veilleux et al. 1995;
Taniguchi et al. 1999; Lutz, Veilleux, \& Genzel 1999). 
This is not, however, unambiguous evidence that there is no AGN present. 
In any case, these measurements do not probe the very central regions,
which is where most of the power is emitted.  Rather they probe regions
with optical depths $\tau < 1$ at optical wavelengths. 

\vskip 1pt \noindent
On the other hand OH megamaser emission certainly indicates that the
power source is compact 
(Lonsdale et al.~1994, 1998). 
This would be consistent with an AGN power source, but again 
this measurement is not a
unique discriminant. 
The nature of the power source in cold ULIGs like Arp 220 has taken on
a particular significance in recent times with the detection of a
population of these objects at high redshifts which contribute a very
large fraction of the extragalactic background light at far-infrared
and submillimeter wavelengths (Barger, Cowie \& Sanders 1999).

We now attempt to provide additional constraints on the nature of the power
source in Arp 220 by investigating the 
so-called ionizing photon deficiency problem.
Prestwich et al.~(1994) performed near-infrared (NIR) spectroscopy for a
sample of starburst galaxies including Arp 220 and found that 
the ratio of
the Br$\gamma$ luminosity to the infrared luminosity of Arp 220
is lower by a factor of ten than what is observed in lower-luminosity 
(but still very luminous)
starburst galaxies, suggesting a deficit of ionizing photons.
This could in part be due to absorption of the ionizing photons by dust,
to an AGN contributing much of $L_{\rm bol}$ but few ionizing photons,
or to the detailed shape of the stellar initial mass function (IMF). 
These physical process all operate on the global properties of Arp 220
in different ways and
in this paper we use data from the literature
together with starburst models and an SED fitting method 
to give constraints 
on the nature of the hidden power source.

\section{MODELS}

The nuclear super-star clusters seen in optical (Shaya et al. 1994) and
near-infrared (Scoville et al. 1998) $HST$ images are not the source
of the high bolometric luminosity of Arp 220 (Shioya, Taniguchi \&
Trentham 2000), which is completely hidden by dust at those
wavelengths.
It is this hidden power source that we attempt to
model.  We use the population synthesis model of Leitherer et al.
(1999, Starburst99) to compute bolometric luminosities and ionizing
photon fluxes which we compare to the observed values derived from
far-infrared and radio data respectively (where extinction is less of
a problem).   

The following assumptions are made:

\noindent
1) we assume a model with continuous star formation.  Note that  
there is abundant cold gas available as fuel 
(Scoville et al. 1991, 1998; Downes \& Solomon 1998); 

\noindent
2) we consider the three stellar IMFs given in Table 1,
where we list the IMF slope $\alpha$, lower-mass cutoff
$M_{\rm l}$, and upper-mass cutoff $M_{\rm u}$.
The values of $M_{\rm l}$ are higher than what is observed in the
solar neighborhood, but seems to be more appropriate
for violent starburst galaxies (e.g., Goldader et al. 1997);

\noindent
3) we assume solar metallicity, appropriate for the central regions of
galaxies, which is where most stars generated in this kind of dense
starburst will end up (e.g.~Faber 1973);  

\noindent
4) we assume that a hypothesized AGN coexists with the
starburst and contributes a fraction
$f_{\rm AGN}$ of the bolometric luminosity.
Radio emission from this AGN is assumed to contribute negligibly to
the ionizing photon flux inferred from radio recombination lines.
This is reasonable since we expect an 
electron column density larger than $\sim 10^{23}{\rm cm}^{-2}$ given the
existence of very dense gas traced by HCN 
(Solomon, Downes, \& Radford 1992) and CS (Solomon,
Radford, \& Downes 1990).
Radio emission from an AGN is then 
suppressed because of free-free absorption.

\section{RESULTS}

\subsection{Initial Mass Function, Age, and AGN fraction}

The total bolometric luminosity of Arp 220 is 
$L_{\rm bol} \simeq 1.5 \times 10^{12} L_{\odot}$ (Sanders et al. 1988). 
When considering the hidden power source, we need to subtract the
part of the bolometric luminosity which {\it is} 
observed at optical and near-infrared wavelengths in the center of
the galaxy. 
The $K$-magnitude of the central $2.5^{\prime \prime}$ diameter region
of Arp 220 is 11.80 mag (Carico et al. 1990). 
Even with $A_V = 6.4$ mag of extinction (Shier et al. 1996), this is only
$1 \times 10^{11}L_{\odot}$ (assuming a bolometric correction of 0.5;
Leitherer et al. 1999). 
This value is less than 10 \% of the total bolometric luminosity of
Arp 220 and we neglect it.
 
The ionizing photon production rate of Arp 220 is
evaluated from the observation of radio recombination lines like  
H92$\alpha$: 
$Q=1.3 \times 10^{55} ~ {\rm photons \; s^{-1}}$
(Anantharamaiah et al. 2000).  This is twice as large as the rate derived 
by Zhao et al. (1996) but Anantharamaiah et al. (2000)
have wider velocity coverage; note that it is also 
larger than the upper limit 
derived by Scoville et al. (1991) from the free--free radio continuum 
flux density at 110 GHz, 
$Q < 7.5 \times 10^{54} ~ {\rm photons \; s^{-1}}$. 
Again, we have to estimate the ionizing photon rate from the regions we
do see in $K$-band and subtract it from the above values to derive the
production rate for the hidden power source.  
The reddening-corrected 
($A_V \simeq$ 10 mag; Larkin et al. 1995) Br$\gamma$ flux
($5.9 \times 10^{-18} ~ {\rm W \; m^{-2}}$; Goldader et al. 1995)
gives an ionizing photon rate, $7 \times 10^{53} ~ {\rm photons \; s^{-1}}$
for these visible starburst regions.  Therefore the production rate for
the hidden power source is: 
$Q({\rm hidden}) 
\simeq 1.23 \times 10^{55} ~ {\rm photons \; s^{-1}}$.

Following Scoville et al. (1997), we adopt the stellar mass 
of the nucleus of Arp 220 as $2.5 \times 10^9M_{\odot}$. 
We therefore set the constraints on the stellar populations 
of the hidden starbursts as follows: 
$8.1 \times 10^{-44} \; L_{\odot}({\rm photons \; s^{-1})^{-1}} < 
L_{\rm bol}/Q({\rm hidden}) < 1.2 \times 10^{-43} \; 
L_{\odot}({\rm photons \; s^{-1})^{-1}}$ and
$400 \; L_{\odot}/M_{\odot} < L_{\rm bol}/M_*({\rm hidden}) < 
600 \; L_{\odot}/M_{\odot}$. 
An AGN fraction $f_{\rm AGN} > 0$ results in both of these parameters
being lowered by a factor $(1-f_{\rm AGN})$. 

\placefigure{fig-1}

Now we compare the above numbers to the parameters derived for 
the three models described in Section 2.
Figure~1 shows the evolutionary locus on a 
$L_{\rm bol}/Q$ - $L_{\rm bol}/M_*$ plane 
for each model.
The general form of the curves is understood as follows.
Low-mass stars have 
smaller $L_{\rm bol}/M_*$ but larger
$L_{\rm bol}/Q$ than high-mass stars.
Since the fractional mass of low-mass stars increases with increasing age,
as time increases,
$L_{\rm bol}/M_*$ becomes smaller while
$L_{\rm bol}/Q$ becomes larger for all the models.

Let us consider the results for the three IMFs in turn.

\subsubsection{IMF1}

For IMF1, the model overpredicts the bolometric luminosity given
the observed ionizing photon flux (box A).  
This is the classical 
ionizing photon deficiency problem (e.g., Goldader et al. 1995). 
This could be solved in a number of ways:
\vskip 1pt \noindent 
(i) reducing the number of massive stars which generate more bolometric
luminosity per unit ionizing photon that lower mass stars.  This could
be achieved by either steepening the slope of the IMF (as in IMF2) or
lowering the upper mass cutoff (as in IMF3).  These adjustments have
different effects on the properties studied here and are discussed in the
next two sections.
\vskip 1pt \noindent 
(ii) if some ionizing photons are absorbed by dust shortly
after their production,
the ionizing photon rate evaluated from 
radio recombination lines will be too low.  The figure shows that
IMF1 can then be consistent with observation if the hidden power source
is a pure starburst with 
$2 \times 10^7 < {\rm age/yr} < 3 \times 10^7$
and between 0.3 and 0.6 of the
ionizing photons that are generated are absorbed by dust 
\vskip 1pt \noindent
(iii) if an AGN contributes some of the power, the ionizing photon
deficiency problem can be reduced. 
If $f_{\rm AGN} = 0.5$, then IMF1 is marginally consistent with
observation if  
$4 \times 10^7 < {\rm age/yr} < 7 \times 10^7$ 
and no more than 30\% of the ionizing photons are absorbed by dust.
Increasing $f_{\rm AGN}$ much beyond 0.5 is not permitted since there
are then too many ionizing photons given the bolometric luminosity and
the ionizing photon deficiency problem now becomes an
ionizing photon {\it excess} problem.  For example, region C on Figure 1
is excluded if IMF1 is valid. 

\subsubsection{IMF2}

This 
steep IMF can only be consistent with observation if the age is very young
($< 5 \times 10^6$ yr) {\it and} $f_{\rm AGN}$ is finely tuned to have a
value of about 0.5 {\it and} if the fraction of ionizing photons absorbed by
dust is finely tuned to have a value of about 0.2 (but not 0).  We therefore
regard it as unlikely.  This IMF can never by itself generate a 
high enough value of $L_{\rm bol}/M_*$ to be consistent with observation.
Increasing the AGN fraction helps a little 
but increasing it by too much creates an ionizing photon excess problem
as described in the previous section. 

\subsubsection{IMF3}

If we adopt the IMF with a smaller upper-mass limit
($M_u=30M_{\odot}$: IMF3), 
the evolutionary loci shift to the upper right on 
Figure 1. 
In the absence of either absorption of ionizing photons by dust or
an AGN, the observational constraints are satisfied if
$6.3 \times 10^6 < {\rm age/yr} < 1.1 \times 10^7$. 
An AGN fraction greater than zero or an appreciable fraction of
ionizing photons being absorbed by dust would generate an
ionizing photon excess problem.

\subsection{Current star formation rate in Arp 220}

Since we adopt the constant star formation model, the
star formation rate (SFR) is coupled uniquely to the bolometric luminosity 
for each model.
Assuming the SFR of $1 \; M_{\odot} \; {\rm yr}^{-1}$,
the ionizing photon production rate is about 
$Q= 4.6 \times 10^{52}~ {\rm photons \; s^{-1}}$ 
at an age of $1 \times 10^7$ yr for Model with IMF3 (i.e.~where it
intersects with Box A on Fig.~1). 
This corresponds to a total SFR of the hidden starburst
of $267 M_{\odot}$  yr$^{-1}$. 
In Arp 220, there are two concentrations of thermal dust emission 
with which the hidden starbursts are associated. 
Downes \& Solomon (1998) evaluated the luminosity ratio between the western 
and the eastern nucleus as 1.5. 
We assume that the ratio of SFR between them is the same as this
luminosity ratio. 
The SFR of the western nucleus is then $160 M_{\odot}$ yr$^{-1}$ and 
that of the eastern nucleus is $107 M_{\odot}$ yr$^{-1}$. 

The results in this and the previous section depend intricately on our
assumption of a star-formation rate that does not vary strongly with time
(Point 1 in Section 2).  An alternative way of explaining the deficiency
is to have a star-formation rate that was extremely high at some time
in the past but is low now.  The highest mass stars, which produce
most of the Lyman continuum, would since have died (these have very short
lifetimes) and therefore the
observed value of $Q$ is low.  If the starburst was sufficiently powerful
to generate many intermediate-mass stars, these could
produce the observed $L_{\rm bol}$.  Very large numbers of such stars would
be required, since each star individually contributes only modestly to
$L_{\rm bol}$.  Such a scenario may then need to be finely tuned in 
terms of both
the time-variation of the star-formation rate and the IMF so as to avoid
overproducing the total mass in stars (this must be less than the dynamical
mass), but this is still an important caveat to our results.  
 
\subsection{Visual Extinction}

Once we have chosen a particular model, we can compute the visual extinction
necessary to hide the hidden starbursts in the $K$-band
$HST$ image of Scoville et al. (1998).
We find that 
$A_V$ must be larger than 30 mag for the western nucleus 
and 40 mag for the eastern nucleus for either IMF1 where more than 0.4 of
the ionizing photons are absorbed by dust or for IMF3 where none of
the ionizing photons are absorbed by dust (see the discussions in Sections
3.1.1 and 3.1.3). 
This result is consistent with the recent MIR observations; i.e.,
$A_V$ of 45 mag (Genzel et al. 1998).

\section{Conclusion}

In all the cases 
that we consider, the hidden power source due to star
formation at the 50\% level or more in order to avoid an ionizing
photon excess problem, and this starburst must be young
($< 7 \times 10^7$ yr).  
We derive a current star formation rate of
$267 M_{\odot}$ yr$^{-1}$, and an extinction $A_V > 30$ mag to this
hidden starburst. 
 
\vspace{1pc}
YS thanks the Japan Society for Promotion of Science (JSPS) 
Research Fellowships for Young Scientist. 
NT thanks the PPARC for financial support.
YT would like to thank R. -P. Kudritzki, Bob McLaren and Dave Sanders
at Institute for Astronomy, University of Hawaii
for their warm hospitality. 
This work was financially supported in part by Grant-in-Aids for the Scientific
Research (Nos.\ 07044054, 10044052, and 10304013) of the Japanese Ministry of
Education, Science, Sports and Culture.


\begin{deluxetable}{cccc}
\footnotesize
\tablecaption{Parameters of IMF
\label{tbl-1}}
\tablewidth{0pt}
\tablehead{
\colhead{IMF} & \colhead{$\alpha$} & \colhead{$M_{\rm u}$} & 
\colhead{$M_{\rm l}$}}
\startdata
1 & 2.35 & 100 & 1 \\
2 & 3.30 & 100 & 1 \\
3 & 2.35 &  30 & 1 \\
\enddata
\end{deluxetable}

\begin{figure}
\epsfysize=17cm \epsfbox{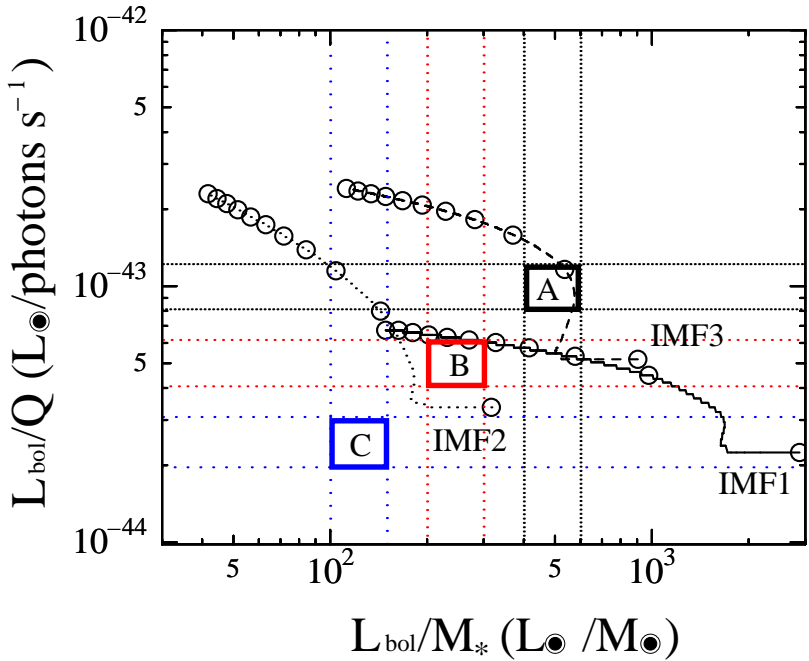}
\caption[]{
The evolutionary locus on the $L_{\rm bol}/Q$
$-$ $L_{\rm bol}/M_*$ plane for the IMF models
listed in Table 1
from $10^5$ yr (the largest $L_{\rm bol}/M_*$ on each locus)
to $10^8$ yr (the smallest $L_{\rm bol}/M_*$ on each locus).
The circles on the line mark intervals of $10^7$ yr.
The black dotted lines show the regions consistent with observation:
$ 8.1 \times 10^{-44}\, L_{\odot} ({\rm photons \; s^{-1}})^{-1} <
L_{\rm bol}/Q <
1.2 \times 10^{-43} \, L_{\odot}
({\rm photons \; s^{-1}})^{-1}$ and
$400 \, L_{\odot}/M_{\odot} < L_{\rm bol}/M_* < 600 \, L_{\odot}/M_{\odot}$.
An AGN fraction $f_{\rm AGN} = 0$ is assumed for the black lines.
The red and blue lines show the region on this diagram consistent with
observation for $f_{\rm AGN} = 0.5$ and 0.75 respectively, assuming that
the AGN contributes only to $L_{\rm bol}$ (see the text for details).
The regions A, B, and C are the regions consistent with observation for
$f_{\rm AGN}$ values of 0, 0.5, and 0.75.
\label{fig1}
}
\end{figure}

\end{document}